\documentclass[onecolumn,showpacs,preprintnumbers,amsmath,amssymb,aps]{revtex4}
\usepackage{graphicx}
\usepackage{amsmath}
\usepackage{slashed}
\usepackage{mathrsfs}
\usepackage{epsfig}
\usepackage{hyperref} 
\usepackage{lipsum}% just to generate text for the example
\usepackage{mathrsfs}
\usepackage{epsfig}
\usepackage{dcolumn}% Align table columns on decimal point
\usepackage{bm}% bold math
\newcommand{\arXhref}[1]{\href{http://arxiv.org/abs/arXiv:#1}{#1}} % For the links to the arXiv preprint
\def\beq{\begin{equation}}
\def\eeq{\end{equation}}
\def\bea{\begin{eqnarray}}
\def\eea{\end{eqnarray}}
\def\nn{\nonumber}
\def\nl{\nonumber \\}

\def\bra{\langle}
\def\ket{\rangle}
\def \nn{\nonumber}
\def \nl{\nn\\}
\def\beq{\begin{equation}}
\def\eeq{\end{equation}}
\def\bea{\begin{eqnarray}}
\def\eea{\end{eqnarray}}
\def\nn{\nonumber}
\def\roughly#1{\mathrel{\raise.3ex\hbox
{$#1$\kern-.75em\lower1ex\hbox{$\sim$}}}}
\def\lsim{\roughly<}
\def\gsim{\roughly>}

\def\sla#1{\raise.15ex\hbox{$/$}\kern-.57em #1}
\def\BDtaunu{\bar{B} \to D^+ \tau^{-} {\bar\nu}_\tau}
\def\BDlnu{\bar{B} \to D^+ \ell^{-} {\bar\nu}_\ell}
\def\BDstartaunu{\bar{B} \to D^{*+} \tau^{-} {\bar\nu}_\tau}
\def\BDstarlnu{\bar{B} \to D^{*+} \ell^{-} {\bar\nu}_\ell}

\def\EPSFIG[#1]#2#3#4{		% Don't be scared by this monsrosity
\begin{figure}[H]		% it is a macro to save typing later
\begin{center}			% 
\includegraphics[#1]{#2}	%
\end{center}			%
\caption{#3}			%
\label{#4}			%
\end{figure}			%
}                              %
\def\scat{ \nu_{\tau}+ n \to \tau + X}

\def\nutau{ \nu_{\tau}}
\def\numu{ \nu_{\mu}}

\def\scat{ \nu_{\tau}+ n \to \tau^- + p}
\def\scatanti{ \bar{\nu}_{\tau}+ p \to \tau^+ + n}

\def\Pmutau{P(\nu_\mu \rightarrow \nu_\tau)}
\def\Pmutaubar{P(\bar{\nu}_\mu \rightarrow \bar{\nu}_\tau)}

\def\mutau{\nu_\mu \rightarrow \nu_\tau}
\def\mutaubar{\bar{\nu}_\mu \rightarrow \bar{\nu}_\tau}
\def\mue{\nu_\mu \rightarrow \nu_e}
\def\muebar{\bar{\nu}_\mu \rightarrow \bar{\nu}_e}

\def\BDtaunu{\bar{B} \to D^+ \tau^{-} {\bar\nu}_\tau}
\def\BDlnu{\bar{B} \to D^+ \ell^{-} {\bar\nu}_\ell}
\def\BDstartaunu{\bar{B} \to D^{*+} \tau^{-} {\bar\nu}_\tau}
\def\BDstarlnu{\bar{B} \to D^{*+} \ell^{-} {\bar\nu}_\ell}

\def\bra#1{\left\langle #1\right|}

\newcommand{\mn}{\Delta m_{21}^2}
\newcommand{\mt}{\Delta m_{31}^2}
\newcommand{\si}{s_{12}}
\newcommand{\sn}{s_{23}}
\newcommand{\st}{s_{13}}
\newcommand{\ci}{c_{12}}
\newcommand{\cn}{c_{23}}

\def\lsim{\raise0.3ex\hbox{$\;<$\kern-0.75em\raise-1.1ex
\hbox{$\sim\;$}}}
\def\gsim{\raise0.3ex\hbox{$\;>$\kern-0.75em\raise-1.1ex
\hbox{$\sim\;$}}}

\begin{document}

\unitlength = 1mm
\begin{flushright}
UMISS-HEP-2016-02 \\
[10mm]
\end{flushright}

\title{Determination of mass hierarchy  with $\nu_\mu \to \nu_\tau$ appearance  and the effect of nonstandard interactions}

\author{Ahmed Rashed$^{1,2,3,4}$\footnote{E-mail:\texttt{amrashed@go.olemiss.edu}}, and Alakabha Datta$^{1}$\footnote{E-mail:\texttt{datta@phy.olemiss.edu}}}
%
%\vspace*{3mm}
\affiliation{
\mbox{$^1$Department of Physics and Astronomy, University of Mississippi,}
\mbox{ Lewis Hall, University, Mississippi, 38677 USA}\\
\mbox{$^2$Department of Physics, Faculty of Science,  Ain Shams University, Cairo, 11566, Egypt}\\
\mbox{$^3$Center for Fundamental Physics, Zewail City of Science and Technology, Giza 12588, Egypt}\\
\mbox{$^4$The Abdus Salam ICTP, Strada Costiera 11, 34014 Trieste, Italy}\\
}

%\newpage
\begin{abstract}
\noindent 
Crucial developments in neutrino physics  would be the determination of the mass hierarchy (MH) and  measurement  of the CP phase in the leptonic sector. The patterns of the transition probabilities $\Pmutau$ and $\Pmutaubar$ are sensitive to these oscillation parameters. An asymmetry parameter can be defined as the difference of these two probabilities normalized to their sum. The profile of the asymmetry parameter gives a clear signal of the mass ordering as it is found to be positive for inverted hierarchy and negative for normal hierarchy. The asymmetry parameter is also sensitive to the CP phase.
We consider the effects of non-standard neutrino interactions (NSI) on the determination of the mass hierarchy. Since we assume the largest new physics effects involve the $\tau$ sector only, we ignore NSI in production and study the NSI effects  in detection as well as along propagation.
% with considering the NSI parameters $(\varepsilon_{\mu\tau},\varepsilon_{\tau\tau})$. 
We find that the NSI effects can significantly modify the prediction of the asymmetry parameter though the MH can still be resolved.
%. Hence the measurement of the asymmetry parameter can be used to determine the mass hierarchy and find evidence for NSI.
% We use the baseline of the Long Baseline Neutrino Oscillation Experiment (LBNO).

%Different techniques have been proposed to be used in oscillation experiments to determine these two parameters. The pattern of the transition probability $P(\nu_\mu \rightarrow \nu_\tau)$ and $P(\bar{\nu}_\mu \rightarrow \bar{\nu}_\tau)$ are sensitive to the mass ordering in the presence of matter effect. Moreover, the comparison of the transition probability of neutrino to anti-neutrino observes an effect of the lepton CP asymmetry. 

\end{abstract}

 \maketitle

% % % % % % % % % % % % % % % % % % % % % % % % % % % % % % % % % % % % % % % % % % % % % % % % % % % % % % % % % % % % % % % % % % % % % % % % % % % % % % % % % % % % % % % % % % % % % % % % % % % % % % % % % % % % % % % % % % % % % % % % % % % % % % % % % % % % % % % % % % 

%\section{Introduction}

After ruling out the zero value of the smallest mixing angle $\theta_{13}$ in the lepton sector with C.L. around $5 \sigma$ \cite{An:2012eh}, the main scope of the future experiments is to answer some open questions such as the absolute mass scale, mass hierarchy, and CP asymmetry in the lepton sector.

Knowledge of the mass hierarchy has an impact on determining the neutrino absolute mass scale, CP asymmetry in the lepton sector, and the nature of the neutrino to be either  Dirac or Majorana.  Once the ordering of the neutrino mass states is determined, the uncertainty on the measurement of the CP-violating phase, $\delta_{\rm CP}$, is significantly reduced. Measuring the mass ordering can cut down the domain for observation of a signal in the neutrinoless double beta decay experiments.  Cosmological measurements are sensitive to the sum of neutrino masses, thus, knowledge of the mass hierarchy could help in determining the absolute neutrino mass scale. 
%For this all, fixing the mass hierarchy represents a step towards understanding more phenomena in neutrino physics. 

The mass hierarchy (MH) can be determined using different techniques. The transition probability from a neutrino flavor to another, in the presence of matter effect,  is sensitive to the mass hierarchy. The shape of the oscillation profile can be used to infer the sign of $\Delta m_{23}^2$ thereby indicating whether we have normal hierarchy (NH) or the inverted hierarchy (IH).
 The standard proposal is to use the appearance channel $\nu_\mu \rightarrow \nu_e$ to measure MH. 
Determination of mass hierarchy is in the scope of several future experiments such as DUNE \cite{Strait:2016mof, Acciarri:2016crz, Acciarri:2016ooe, Acciarri:2015uup}, Hyper-Kamiokande \cite{Abe:2014oxa, Abe:2011ts}, LBNO \cite{::2013kaa, Agarwalla:2014tca}, and INO \cite{Athar:2006yb}. A variety of other experiments has some sensitivity to the mass hierarchy such as the reactor neutrino experiments JUNO (formerly known as Daya Bay II) and RENO as well as PINGU \cite{Akhmedov:2012ah} at IceCube. The CP asymmetry can be measured in very long base line neutrino experiments such as LBNO (2300 km baseline length) and DUNE (1300 km baseline length) as well as Hyper-K (295 km baseline length).
The existence of neutrino masses and mixing require physics beyond the standard model (SM). Hence it is not unexpected that neutrinos could  have non-standard interactions (NSI).  An important question is how this NSI impact the MH determination or the measurement of the CP violating phase \cite{NSIref}. Even if NSI does not significantly impact  the
MH determination it will be useful to have alternate channels to confirm the results from the standard channel.

In this work we want to explore MH in the $\nu_\mu \rightarrow \nu_\tau$ channel.
  Compared to the standard channel, $\nu_\mu \rightarrow \nu_\tau$ has certain advantages. The transition probability for $\nu_\mu \rightarrow \nu_\tau$ is proportional to sine of the  atmospheric mixing angle, while $P(\nu_\mu \rightarrow \nu_e)$ is suppressed by small oscillation parameters, such as $\sin^2 \theta_{13}$ and $\Delta m_{12}^2$. 
  We are going to focus, in this paper, on  the long baseline DUNE and LBNO experiments. We will also consider NSI effects in our analysis.

There are several reasons to consider NSI involving the $( \nu_\tau, \tau)$ sector. First, the third generation may be more sensitive to new physics effects because of their larger masses. As an example, in certain versions of the two Higgs doublet models (2HDM) the couplings of the new Higgs bosons are proportional to the masses and so new physics effects are more pronounced for the third generation. Second, the constraints on new physics (NP) involving the third generation leptons are somewhat weaker allowing for larger new physics effects.

A key property of the SM gauge interactions is that they are lepton flavor universal. Evidence for violation of this property would be a clear sign of new physics (NP) beyond the SM. 
Interestingly, there have been some reports of non universality in the lepton sector from experiments.  In the charged current sector the decays $\bar{B} \to D^{(*)+}
\ell^{-} {\bar\nu}_\ell$,  have been measured by the BaBar
\cite{RD_BaBar}, Belle \cite{RD_Belle} and LHCb \cite{RD_LHCb}
Collaborations. It is found that the values of the ratios ${\cal
  B}(\bar{B} \to D^{(*)+} \tau^{-} {\bar\nu}_\tau)/{\cal B}(\bar{B}
\to D^{(*)+} \ell^{-} {\bar\nu}_\ell)$ ($\ell = e,\mu$) deviate from the SM predictions \cite{RDtheory} and this could be indication of lepton non universal new physics\cite{datta}.
Specifically, \cite{RDRK_Isidori}
\bea
R_D &\equiv& \frac{ {\cal B}(\BDtaunu)_{expt} / {\cal B}(\BDtaunu)_{SM} }
{ {\cal B}(\BDlnu)_{expt} / {\cal B}(\BDlnu)_{SM} }  =  1.37 \pm 0.18 ~, \nonumber\\
R_{D^*} &\equiv& \frac{ { B}(\BDstartaunu)_{expt} / { B}(\BDstartaunu)_{SM} }
{ { B}(\BDstarlnu)_{expt} / { B}(\BDstarlnu)_{SM} } =  1.28 \pm 0.08 ~.
\label{indirect2}
\eea
The measured values of $R_D$ and $R_{D^*}$ represent deviations from
the SM of 2.0$\sigma$ and 3.8$\sigma$, respectively. There also appears to be violation of lepton universality in $W \tau \nu_\tau$ coupling though it is difficult to explain \cite{Wtaun}.

There has been another recent hint of lepton non-universality in the neutral current sector. The
LHCb Collaboration measured the ratio of decay rates for $B^+ \to K^+
\ell^+ \ell^-$ ($\ell = e,\mu$) in the dilepton invariant mass-squared
range 1 GeV$^2$ $\le q^2 \le 6$ GeV$^2$ \cite{RKexpt}, and found
\bea
R_K & \equiv & \frac{{\cal B}(B^+ \to K^+ \mu^+ \mu^-)}{{\cal B}(B^+ \to
  K^+ e^+ e^-)} \nn\\
&=& 0.745^{+0.090}_{-0.074}~{\rm (stat)} \pm 0.036~{\rm (syst)} ~.
\eea
This differs from the SM prediction of $R_K = 1 \pm O(10^{-4})$
\cite{RKtheory} by $2.6\sigma$. 

These measurements  might be hinting towards lepton non universal new physics with the largest effects involving the third generation leptons \cite{GGL}. The new physics could arise in  the third generation and feed down to other generation through mixing effects and so in this picture we expect the largest NSI to involve the third generation neutrino.
In our analysis, therefore, we will assume NSI only involving the third generation leptons.

%This appearance channel is relevant for experiments like Hyper-K and OPERA \cite{OPERA:2011oja}. Super-Kamiokande experiment searched \cite{Abe:2012jj} for the appearance of tau neutrinos from the oscillation of muon neutrinos generated in the atmosphere. They look for tau leptons resulting from the interactions of oscillation-generated tau neutrinos in the detector. They estimated that $180.1\pm 44.3 (stat)^{+17.8}_{-15.2} (syst)$ tau leptons were produced in the 22.5 kton fiducial volume of the detector by tau neutrinos during the 2806 day running period.

The tau-neutrino appearance channel is relevant to the Long Baseline Neutrino Oscillation Experiment (LBNO) which has an access to both transitions $\mutau$ and $\mutaubar$. The experiment consists of a near detector at CERN in addition to a far detector situated at Pyh\"asalmi in Finland 2300 km away from CERN, where the source of neutrino beam is located. 
%The main goal of the experiment is to search for mass hierarchy and CP violating phase in the appearance channels $\mue$ and $\muebar$. It is %proposed to determine the MH to $> 5 \sigma$ C.L. in 5 years of running.
 The muon- neutrino and anti-neutrino fluxes fall in the energy range of $0-10$ GeV where it peaks at 5 GeV \cite{::2013kaa}. This means that the quasi-elastic neutrino interaction is dominant in the energy range of the experiment. 

 An upcoming experiment is the Deep Underground Neutrino Experiment (DUNE) experiment which has a program to make precise measurements of the mixing between the neutrinos, CP violation, and the ordering of neutrino masses. The two main oscillation channels are $\mue$ and $\muebar$, but access to $\mutau$ and $\mutaubar$ modes is possible. The baseline of DUNE is 1300 km and the flux of the neutrino beam ranges from 0-10 GeV \cite{Acciarri:2015uup}.

In table 3 in Ref.~\cite{::2013kaa}, one can find a comparison between LBNO and DUNE. In the case of LBNO, the expected number of events in the channel $\mutau$ that comes from charged current interactions is 215/239 for NH/IH, while the number for $\mutaubar$ is anticipated to be 98/99 for NH/IH in 2.5 years of data-taking. 
%These numbers of events are appropriate to construct the pattern of the transition probability.
 The DUNE  will observe  less number of events in these channels. 
 %The disappearance channels ($\nu_\mu \rightarrow \nu_\mu$ and $\bar{\nu}_\mu \rightarrow \bar{\nu}_\mu$)  independently determine the $\nu_%\mu / \bar{\nu}_\mu$ fluxes at the far detector. 

%In this paper we study the determination of the mass hierarchy and CP violation in the $\nu_\mu \to \nu_\tau$ oscillation. 
The  pattern of the transition probability of $\nu_\mu \to \nu_\tau$ depends on the sign of $\Delta m_{23}^2$ and the CP violating phase $\delta$, so one can extract information on MH and the CP phase from this probability. An asymmetry parameter $A$ can be defined as the difference of the two transition probabilities $\Pmutau$ and $\Pmutaubar$ normalized to their sum.  The MH can be sensitive to the sign of the asymmetry and the size of the asymmetry can carry some sensitivity to the CP violating phase $\delta$.
%One can find that in the scenario of normal/inverted hierarchy, $A$ found to be negative/positive in the LBNO energy %range. This shows a very clear technique in determining the mass ordering. Also, a hint of the CP asymmetry can be %extracted as the pattern of $A$ is sensitive to $\delta$.

We also consider the NSI effects on the determination of MH. 
The NSI effects can arise at the source, along propagation and at detection. Assuming significant NSI only involving the third generation we will ignore NSI at the source.
In any new physics model NSI in propagation and detection are connected. However, we will not use specific models and instead will adopt an effective Lagrangian description of the new physics effects. The NSI effects are parameterized by some co-efficients that depend on the parameters in the effective Lagrangian and we will use experiments to constrain the size of these effects.
Along propagation we will consider the effects of the NSI parameters $(\varepsilon_{\mu\tau},\varepsilon_{\tau\tau})$ to the transition probability pattern (\cite{Adamson:2013ovz, Mitsuka:2011ty, Choubey:2015xha}).  For simplicity we will assume the NSI parameters to be real.
Discussion on the impact of NSI parameters (moduli and phases) for CP violation measurement using $P(\nu_\mu \rightarrow\nu_e)$ for the DUNE experiment can be found in Ref.~\cite{Masud:2016bvp}. %to $ to study the effects on atmospheric mixing angle measurements and to probe the lepton non-universality in tau neutrino scattering.

In previous work \cite{Rashed:2012bd, Rashed:2013dba, Liu:2015rqa} we introduced NSI at detection
and considered various phenomenology connected  to neutrino physics.
For NSI at detection we use the following picture.
The measurement of the transition probability $P(\nu_\mu \rightarrow \nu_\tau)$ can be expressed as \cite{relationship}:
\beq
N(\nu_\tau) = P(\nu_\mu \rightarrow \nu_\tau) \times \Phi (\nu_\mu)\times \sigma_{{\rm SM}}(\nu_\tau)\,,
\label{eq-1}
\eeq
where $N(\nu_\tau)$ is the number of observed events, $\Phi (\nu_\mu)$ is the flux of muon neutrinos at the detector,   $\sigma_{{\rm SM}}(\nu_\tau)$  is the total cross section of tau neutrino interactions with nucleons in the SM at the detector, and $P(\nu_\mu \rightarrow \nu_\tau)$ is the probability for the flavor transition $\numu \to \nutau$ in the presence of matter effect. 
In the presence of  NSI at the detector, Eq.~\ref{eq-1} is modified as
\beq
N (\nu_\tau) = P_{{\rm tot}}(\nu_\mu \rightarrow \nu_\tau) \times \Phi (\nu_\mu)\times \sigma_{{\rm tot}}(\nu_\tau),
\label{eq-2}
\eeq
with $\sigma_{{\rm tot}}(\nu_\tau)=\sigma_{{\rm SM}}(\nu_\tau)+\sigma_{{\rm NP}}(\nu_\tau)$, where $\sigma_{{\rm NP}}(\nu_\tau)$ refers to the additional terms to the SM contribution towards the total cross section. Hence, $\sigma_{{\rm NP}}(\nu_\tau)$  includes  contributions from both the SM and NP interference amplitudes, and the pure NP amplitude. { }From Eqs.~(\ref{eq-1}, \ref{eq-2})
\beq
P_{{\rm tot}}(\nu_\mu \rightarrow \nu_\tau) = P(\nu_\mu \rightarrow \nu_\tau) \frac{\sigma_{{\rm SM}}(\nu_\tau)}{\sigma_{{\rm tot}}(\nu_\tau)}.
\eeq
%The $\nu_\mu / \bar{\nu}_\mu$ fluxes are determined in LBNO by the disappearance channels $(\nu_\mu\rightarrow%%\nu_\mu)$ and $(\bar{\nu}_\mu\rightarrow\bar{\nu}_\mu)$, note that we ignore NP contribution to the muon-neutrino %interactions at the near and far detectors.

Moving on to the transition probabilities, we define the asymmetry parameter as the difference between the neutrino and anti-neutrino transition probabilities normalized to their sum
\beq
A = \frac{P_{{\rm tot}}(\nu_\mu \rightarrow \nu_\tau) - P_{{\rm tot}}(\bar{\nu}_\mu \rightarrow \bar{\nu}_\tau)}{P_{{\rm tot}}(\nu_\mu \rightarrow \nu_\tau) + P_{{\rm tot}}(\bar{\nu}_\mu \rightarrow \bar{\nu}_\tau)}.
\eeq
%We study the effect of the contributions of NP models to $A$. 
In the limit where matter effects are neglected $A$ is just a measure of CP violation.

The transition probability of the appearance channel $\nu_{\mu} - \nu_{\tau}$ in the presence of matter effect and NSI along propagation is given as \cite{Kikuchi:2008vq, Meloni:2009ia, Ohlsson:2012kf, Agarwalla:2015cta}
%To present the oscillation probabilities in the $\nu_{\mu} - \nu_{\tau}$ sector,
%we start by recapitulating the decomposition formula 
%(\ref{Pmutau-decomposition}) in Sec.~\ref{mu-tau-sector}: 
%
\begin{eqnarray} 
P(\nu _\alpha \rightarrow \nu _\beta; 
\varepsilon_{e \mu }, \varepsilon_{e \tau }, \varepsilon_{\mu \mu }, \varepsilon_{\mu \tau }, \varepsilon_{\tau \tau }) &=& 
P(\nu _\alpha \rightarrow \nu _\beta; \text{2 flavor in vacuum}) 
\nonumber \\
&+& 
P(\nu _\alpha \rightarrow \nu _\beta; 
\varepsilon_{e \mu }, \varepsilon_{e \tau }) 
\nonumber \\
&+& 
P(\nu _\alpha \rightarrow \nu _\beta; 
\varepsilon_{\mu \mu }, \varepsilon_{\mu \tau }, \varepsilon_{\tau \tau }) ,
\label{Pmutau-decomposition2}
\end{eqnarray}
where $\alpha$ and $\beta$ denote one of $\mu$ and $\tau$, and $\varepsilon$'s are the NSI parameters. 
The first term in Eq.~\ref{Pmutau-decomposition2} has a form that it appears 
in the two flavor oscillation in vacuum: 
%
%%%%%%%%%%%%%%% P_MUTAU VAC %%%%%%%%%%%%%%%%
\begin{eqnarray}
P(\nu _\mu \rightarrow \nu _\tau; \text{2 flavor in vacuum}) &=& 
4\cn ^2\sn ^2\sin ^2\frac{\mt L}{4E},
\label{Pmutau-vac}
\end{eqnarray}
%%%%%%%%%%%%%%% P_MUTAU VAC %%%%%%%%%%%%%%%%
where $s_{ij}\equiv \sin \theta_{ij}$ and $c_{ij}\equiv \cos \theta_{ij}$. 
The second and third terms in the oscillation probability in the 
$\nu_{\mu} \rightarrow \nu_{\tau}$ channel are given by 
%
%%%%%%%%%%%%%%% P_MUTAU %%%%%%%%%%%%%%%%
\begin{eqnarray}
&& \hspace{-5mm}
P(\nu _\mu \rightarrow \nu _\tau; \varepsilon_{e \mu }, \varepsilon_{e \tau }) 
\nonumber \\
&=& 4 \cn ^2\sn ^2 \vert \Xi \vert^2 
\left( \frac{aL}{4E} \right) \sin \frac{\mt L}{2E} 
- 8\cn ^2\sn ^2 \vert \Xi \vert^2 
\sin \frac{aL}{4E}\sin \frac{\mt L}{4E}\cos \frac{\mt -a}{4E}L 
\nonumber \\
&+& 4\cn ^2\sn ^2 \vert \Theta_{\pm} \vert^2 
\left( \frac{a}{\mt -a} \right) \left( \frac{aL}{4E} \right) \sin \frac{\mt L}{2E} 
\nonumber \\
&-& 8\cn ^2\sn ^2 \vert \Theta_{\pm} \vert^2 
\biggl (\frac{a}{\mt -a}\biggr )^2\cos \frac{aL}{4E}\sin \frac{\mt L}{4E}\sin \frac{\mt -a}{4E}L 
\nonumber \\
&+& 8\cn \sn (\cn ^2-\sn ^2) \vert \Xi \vert  \vert \Theta_{\pm} \vert \cos (\xi - \theta_{\pm}) 
\left( \frac{a}{\mt -a} \right) 
\left ( \frac{a}{\mt} \right) \sin ^2\frac{\mt L}{4E} 
\nonumber \\
&+& 8\cn \sn \vert \Xi \vert  \vert \Theta_{\pm} \vert 
\left( \frac{a}{\mt -a} \right) 
\sin \frac{aL}{4E}\sin \frac{\mt L}{4E} 
\nonumber \\
&& \hspace{14mm}  \times 
\left[ 
\sn ^2 \cos \left( \xi - \theta_{\pm} - \frac{\mt -a}{4E}L \right)  
- \cn ^2 \cos \left( \xi - \theta_{\pm} + \frac{\mt -a}{4E}L \right)  
\right], 
\label{Pmutau-emu-etau}
\end{eqnarray}
and

%%%%%%%%%%%%%%% P_MUMU %%%%%%%%%%%%%%%%
\begin{eqnarray}
&& \hspace{-5mm}
 P(\nu_{\mu} \rightarrow \nu_{\tau}; \varepsilon_{\mu \mu }, \varepsilon_{\mu \tau }, \varepsilon_{\tau \tau })\nonumber\\ 
 &=&- 2\cn ^2\sn ^2 
\left(  \st ^2\frac{\mt}{a} - \mathcal{S}_{1} \right) 
\left( \frac{aL}{2E} \right) \sin \frac{\mt L}{2E} 
+ \cn ^2\sn ^2 
\mathcal{S}_{1}^2
\biggl (\frac{aL}{2E}\biggr )^2\cos \frac{\mt L}{2E} 
\nonumber \\
&-& 8\cn \sn (\cn ^2-\sn ^2) 
\left [ 
\ci \si \st \cos \delta \left( \frac{\mn}{a} \right)  - \vert \mathcal{E} \vert \cos \phi  
\right ] 
\left( \frac{a}{\mt} \right) \sin ^2\frac{\mt L}{4E} 
\nonumber \\
%%%%%%%%%%%
&+& 4\cn \sn (\cn ^2-\sn ^2)  \mathcal{S}_{1} 
\vert \mathcal{E} \vert \cos \phi 
 \biggl (\frac{a}{\mt}\biggr )
\left[ 
\left( \frac{aL}{2E} \right) \sin \frac{\mt L}{2E}-2 \biggl (\frac{a}{\mt}\biggr ) \sin ^2\frac{\mt L}{4E} 
\right] 
\nonumber \\
&+& 4\cn ^2\sn ^2 
\vert \mathcal{E} \vert^2
\left( \frac{a}{\mt}\frac{aL}{2E} \right) \sin \frac{\mt L}{2E} 
\nonumber \\
&+& 4 \vert \mathcal{E} \vert^2 
\biggl [ 
(\cn ^2-\sn ^2)^2 - 4 \cn ^2\sn ^2 \cos^2 \phi 
\biggr ] 
\biggl (\frac{a}{\mt}\biggr )^2\sin ^2\frac{\mt L}{4E}. 
\label{Pmumu-mu-tau}
\end{eqnarray}
%%%%%%%%%%%%%%% P_MUMU %%%%%%%%%%%%%%%%
%

%
The subscript $\pm$ in these equations 
denote the normal and the inverted mass hierarchies, which corresponds 
to the positive and negative values of $\Delta m^2_{32}$.
The simplified notations which involve $\varepsilon$'s in the 
$\nu_\mu - \nu_\tau$ sector 
%as well as $\varepsilon_{ee}$.\footnote{
%%%%%%%%%%%%%%%%% footnote %%%%%%%%%%%%%%%%%
%For readers who want to see the fully explicit expressions of all the oscillation probabilities, we refer the first arXiv version of this paper \cite{KMU-arXiv-v1}. }
%%%%%%%%%%%%%%%%% footnote %%%%%%%%%%%%%%%%%
%%
%Together with the ones already defined in Sec.~\ref{nu_e-sector}, 
%they 
are as follows: 
\begin{eqnarray} 
\Theta_{\pm} &\equiv& 
\st \frac{\mt}{a} + (s_{23} \varepsilon_{e \mu} + c_{23}  \varepsilon_{e \tau} ) e^{i \delta} 
\equiv \vert \Theta_{\pm} \vert e^{i \theta_{\pm} }, 
\nonumber \\ 
\Xi &\equiv& 
\left( \ci \si \frac{\mn}{a}+ c_{23} \varepsilon_{e \mu} - s_{23} \varepsilon_{e \tau} \right) e^{i \delta} 
\equiv \vert \Xi \vert e^{i \xi}, 
\nonumber \\
\mathcal{E} &\equiv& 
\cn \sn (\varepsilon _{\mu \mu}-\varepsilon _{\tau \tau})+\cn ^2\varepsilon _{\mu \tau}-\sn ^2\varepsilon _{\mu \tau}^* 
\equiv \vert \mathcal{E} \vert e^{i \phi},
\nonumber \\
\mathcal{S}_{1} &\equiv& 
 (\cn ^2-\sn ^2)(\varepsilon _{\tau \tau}-\varepsilon _{\mu \mu})+2\cn \sn (\varepsilon _{\mu \tau}+\varepsilon _{\mu \tau}^*)-\ci ^2\frac{\mn}{a}. 
\label{def-recapit}
\end{eqnarray}
We also note that $\Theta_{\pm}$, $\Xi$, and $\mathcal{E}$ are 
complex numbers while $\mathcal{S}_{1}$ is real. The matter potential is given by 
\begin{eqnarray}
a&=&2\sqrt{2}G_F N_e E  \nonumber\\
&=&7.6324 \times 10^{-5} {\rm eV}^2 
\frac{\rho}{{\rm gcm^{-3}}} \frac{E}{{\rm GeV}}.
\end{eqnarray}
%and
%\begin{eqnarray}
%\Delta_{ij}^{\prime} &\equiv &\frac{\Delta_{ij}L}{4E}
%\equiv \frac{(m_i^2-m_j^2) L}{4E}.
%%a^{\prime}&\equiv &\frac{a L}{4E}
%%\equiv \frac{(m_i^2-m_j^2) L}{4E}. 
%\label{3}
%\end{eqnarray}
Using the Preliminary Reference Earth Model (PREM) \cite{PREM:1981}, 
the line-averaged constant matter density is $\rho=3.54\,\mathrm{g/cm^3}$
for the LBNO baseline of $L=2300\,\mathrm{km}$ which 
corresponds to the distance between CERN and Pyh\"asalmi  \cite{Agarwalla:2011hh,Stahl:2012exa,Agarwalla:2014tca}. For the DUNE experiment, we use the standard value of the matter density $\rho=2.8\,\mathrm{g/cm^3}$.
In matter, the probability for T conjugate channels is obtained by the replacement $\delta \to -\delta$ and those for CP conjugate channels are obtained by $\delta \to -\delta$ and $a \to -a$.

\begin{figure*}
\centering
 \includegraphics[width=7cm]{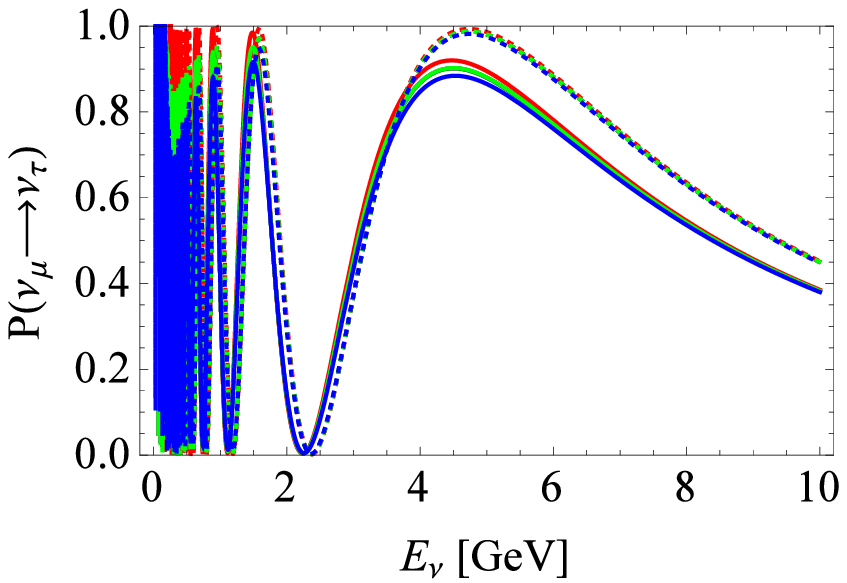}~~~
 \includegraphics[width=7cm]{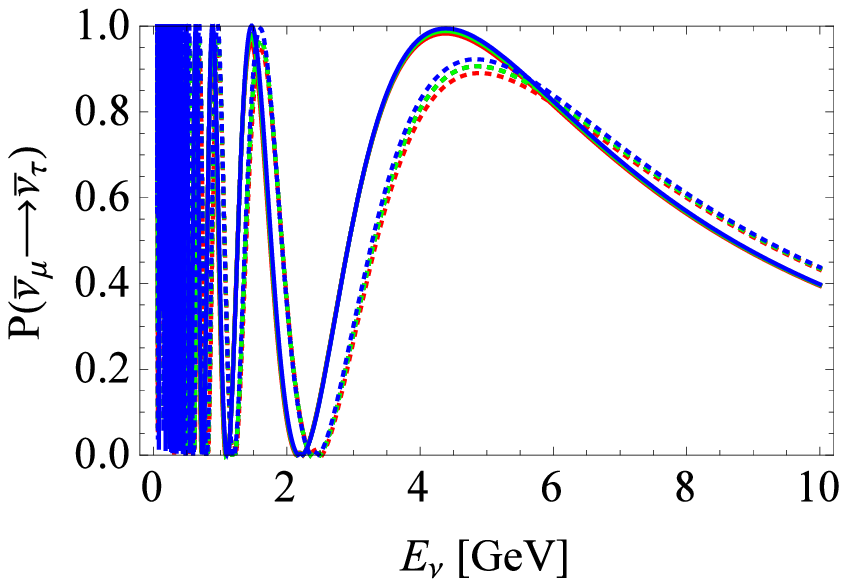}~~~\\
  \includegraphics[width=7cm]{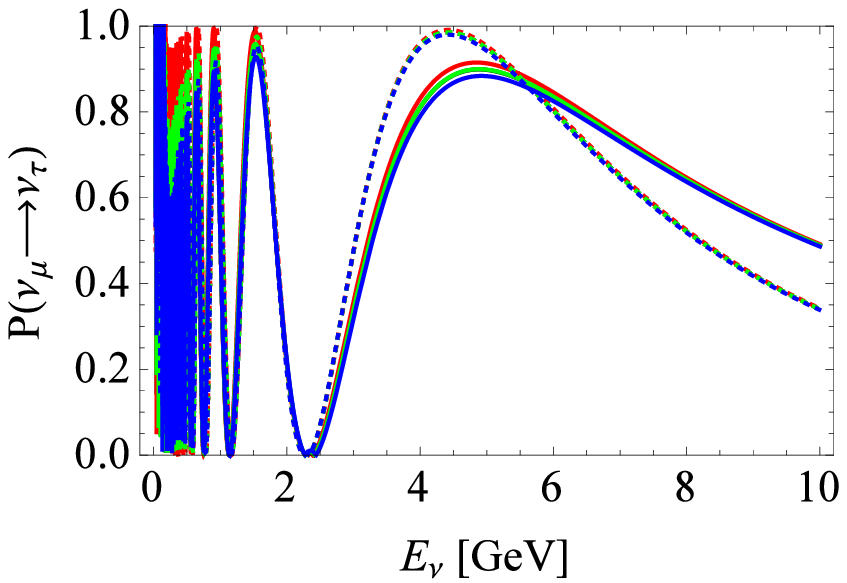}~~~
 \includegraphics[width=7cm]{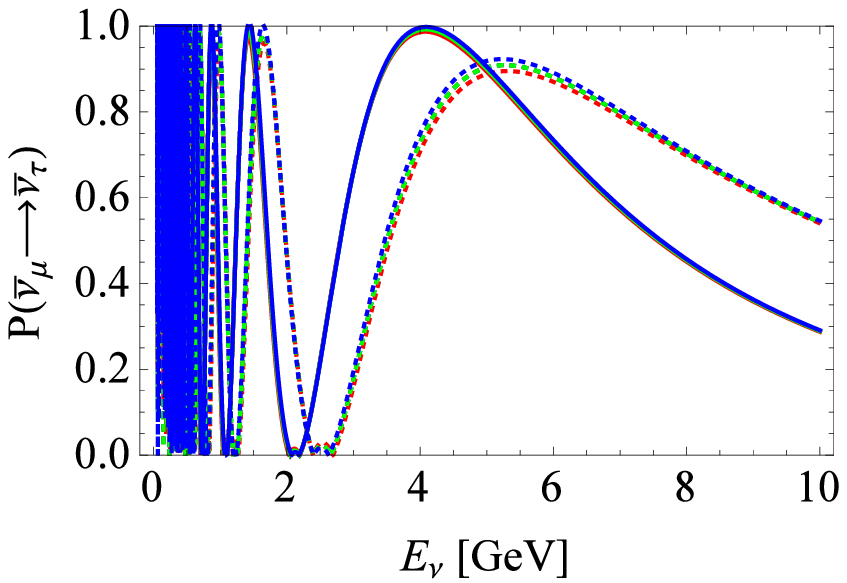}~~~
\caption{The transition probability of the $\nu_\mu \rightarrow \nu_\tau$ (left) channel and its CP conjugate channel $\bar{\nu}_\mu \rightarrow \bar{\nu}_\tau$ (right) in the presence of matter effect for the LBNO energy range and baseline. The solid/dotted lines correspond to NH/IH. The green, red, and blue lines correspond to $\delta=(0,\pi/2,-\pi/2)$, respectively. NSI parameters are taken to be $(\varepsilon_{\mu\tau},\varepsilon_{\tau\tau})=(0,0)$ (top) and $(\varepsilon_{\mu\tau},\varepsilon_{\tau\tau})=(0.07,0.147)$ (bottom).}
\label{Figure1-1}
\end{figure*}

\begin{figure*}
\centering
\includegraphics[width=7cm]{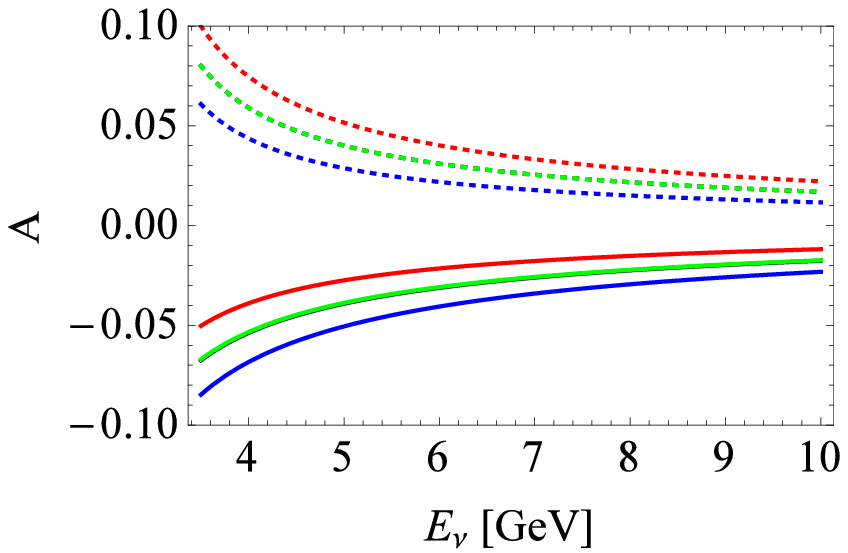}~~~
\includegraphics[width=7cm]{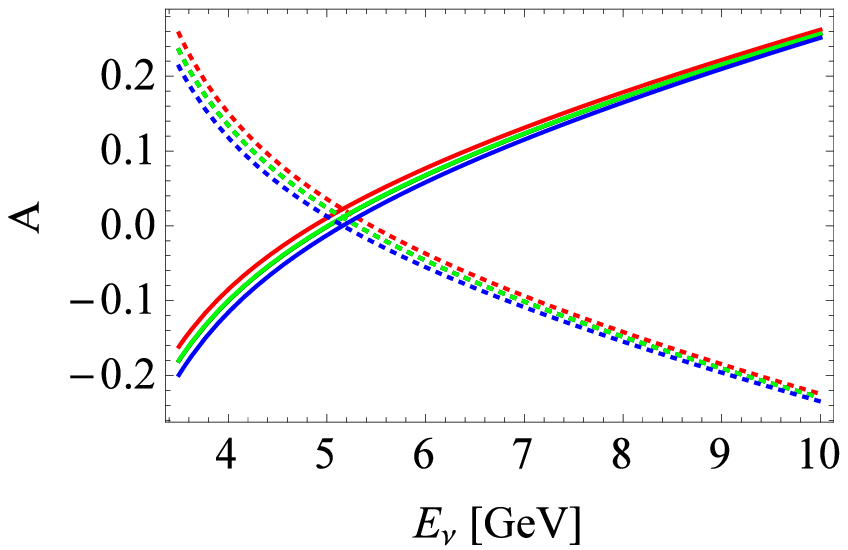}~~~
\caption{The asymmetry parameter $A$  for the LBNO energy range and baseline length. The solid/dotted lines correspond to NH/IH. The green, red,  and blue lines correspond to $\delta=(0,\pi/2,-\pi/2)$, respectively. NSI parameters are taken to be $(\varepsilon_{\mu\tau},\varepsilon_{\tau\tau})=(0,0)$ (left) and $(\varepsilon_{\mu\tau},\varepsilon_{\tau\tau})=(0.07,0.147)$ (right).}
\label{Figure2-1}
\end{figure*}

\begin{figure*}
\centering
 \includegraphics[width=7cm]{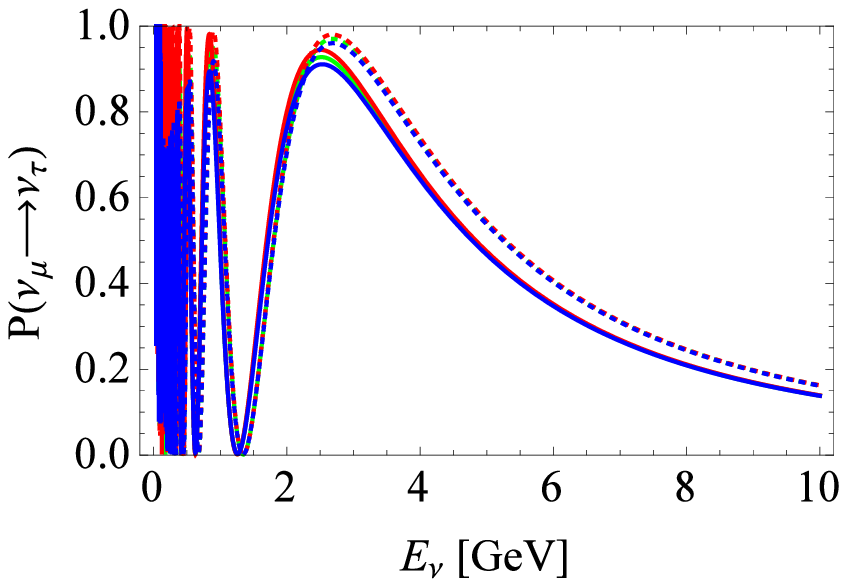}~~~
 \includegraphics[width=7cm]{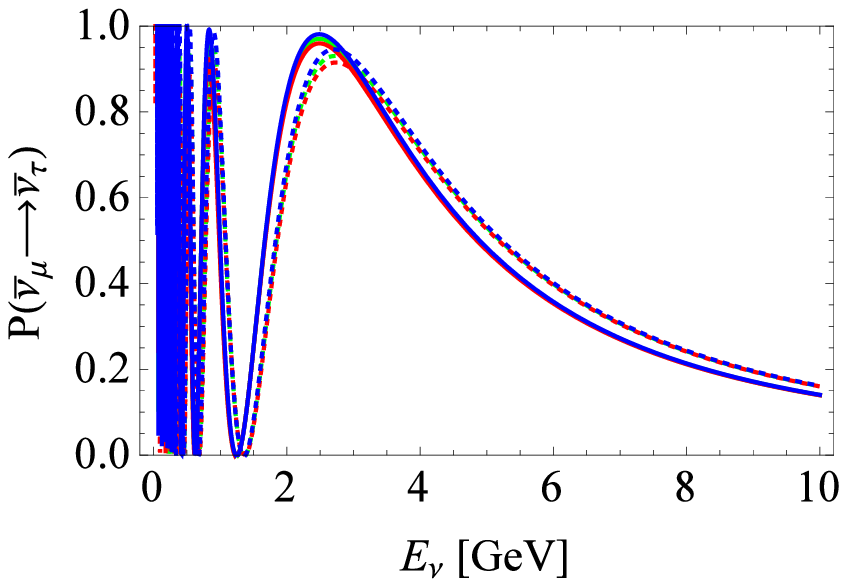}~~~\\
  \includegraphics[width=7cm]{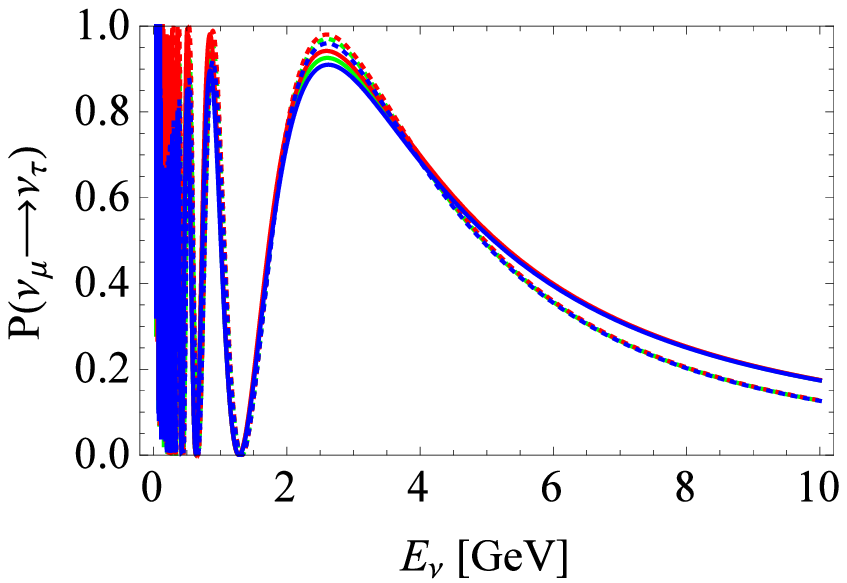}~~~
 \includegraphics[width=7cm]{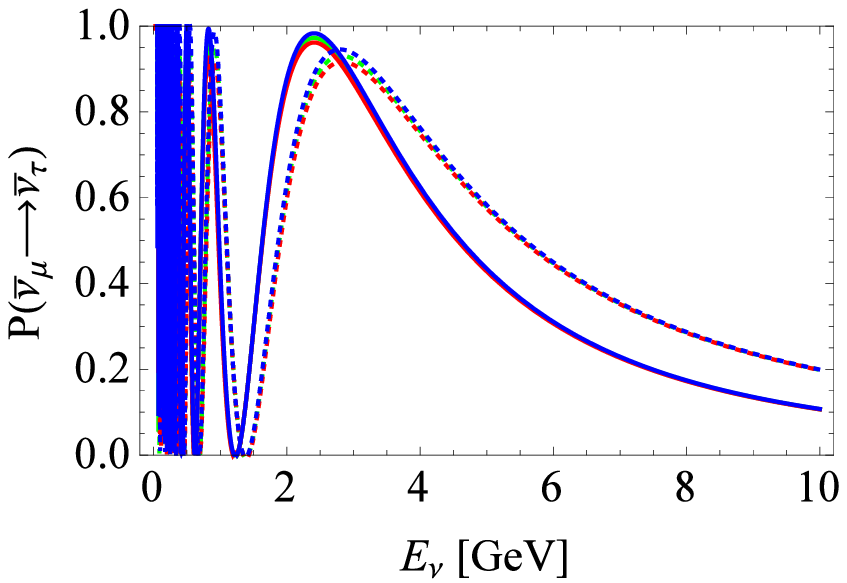}~~~
\caption{The transition probability of the $\nu_\mu \rightarrow \nu_\tau$ (left) channel and its CP conjugate channel $\bar{\nu}_\mu \rightarrow \bar{\nu}_\tau$ (right) in the presence of matter effect for the DUNE energy range and baseline. The solid/dotted lines correspond to NH/IH. The green, red, and blue lines correspond to $\delta=(0,\pi/2,-\pi/2)$, respectively. NSI parameters are taken to be $(\varepsilon_{\mu\tau},\varepsilon_{\tau\tau})=(0,0)$ (top) and $(\varepsilon_{\mu\tau},\varepsilon_{\tau\tau})=(0.07,0.147)$ (bottom).}
\label{Figure1-1DUNE}
\end{figure*}

\begin{figure*}
\centering
\includegraphics[width=7cm]{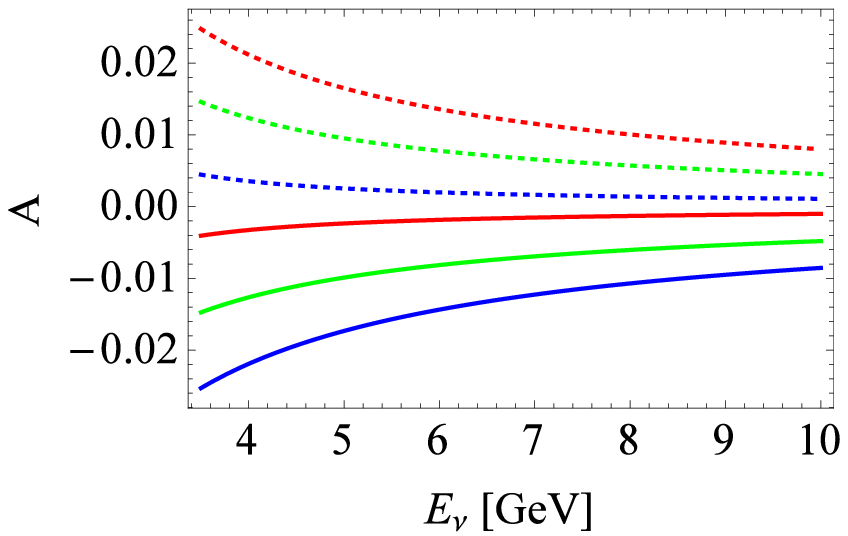}~~~
\includegraphics[width=7cm]{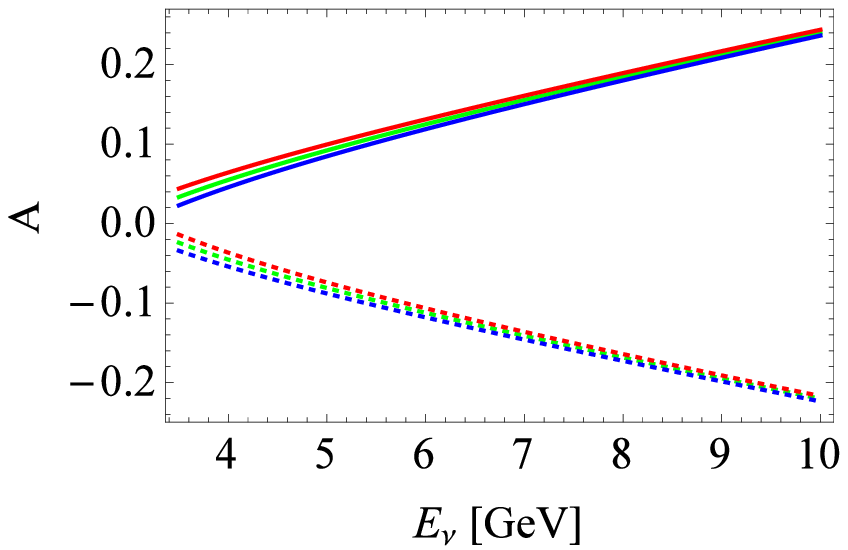}~~~
\caption{The asymmetry parameter $A$  for the DUNE energy range and baseline length. The solid/dotted lines correspond to NH/IH. The green, red,  and blue lines correspond to $\delta=(0,\pi/2,-\pi/2)$, respectively. NSI parameters are taken to be $(\varepsilon_{\mu\tau},\varepsilon_{\tau\tau})=(0,0)$ (left) and $(\varepsilon_{\mu\tau},\varepsilon_{\tau\tau})=(0.07,0.147)$ (right).}
\label{Figure2-1DUNE}
\end{figure*}

In Fig.~\ref{Figure1-1} we show $P(\nu_\mu \rightarrow \nu_\tau)$ and its CP conjugate channel in the LBNO energy range. Here, we consider no  NSI  along propagation  $(\varepsilon_{\mu\tau},\varepsilon_{\tau\tau})=(0,0)$ (top panel) and with  experimental upper bound of $(\varepsilon_{\mu\tau},\varepsilon_{\tau\tau})=(0.07,0.147)$ (bottom panel) (\cite{Adamson:2013ovz, Mitsuka:2011ty, Choubey:2015xha}). Other NSI parameters are taken to be zero. 
In Fig.~\ref{Figure2-1}, we show the asymmetry parameter $A$ which is positive for IH and decreases with energy, while it is negative for NH and increases with energy  for $(\varepsilon_{\mu\tau},\varepsilon_{\tau\tau})=(0,0)$. In the presence of NSI $(\varepsilon_{\mu\tau},\varepsilon_{\tau\tau})=(0.07,0.147)$, $A$ changes sign for both the hierarchies. The asymmetry profile for the two hierarchies has a crossing point where the MH cannot be resolved. The shape of the $A$ parameter in this case can therefore  resolve the MH ( except at the crossing point)  and provides clear evidence of NSI.
One can  notice that that $A$ parameter is sensitive to the CP  phase. The same plots for the energy range and baseline relevant to DUNE experiment are shown in Figs.~(\ref{Figure1-1DUNE}, \ref{Figure2-1DUNE}). Compared to LBNO results, one can find that the asymmetry parameter has smaller values  with considering no NSI along propagation and so it will be difficult to resolve the MH. In this case, $A$,
  does not flip sign, in the desired energy range, when NSI along propagation is included. However, at larger energies in the presence of NSI, $A$ is substantially different for the two hierarchies so that the MH can be resolved.

%The transition probability of $\nu_\mu \rightarrow \nu_\tau$ is bigger in value than $\nu_\mu \rightarrow \nu_e$ as shown in Figs.~(\ref{Figure1-1}, \ref{Figure1-11}). Figure \ref{Figure1-11} matehes the plots in Ref.~\cite{::2013kaa}. This means that the measurement of MH in the former channel is more sensitive. 

Now, we study the effects of new physics contributions to the tau-neutrino  interactions at the detector on the pattern of the asymmetry parameter $A$ in the energy range relevant to LBNO and DUNE. Adopting an effective Hamiltonian approach we include   generic vector axial-vector, scalar, and tensor interactions. 
%Since,  in many realistic
%models both the scalar and tensor couplings may be present,  we consider an explicit leptoquark model where both the scalar %and tensor couplings are present.
 In the LBNO and DUNE energy range, the quasielastic tau neutrino scattering is dominant. 
%NSI parameters $(\varepsilon_{\mu\tau},\varepsilon_{\tau\tau})$ along propagation are considered.

% % % % % % % % % % % % % % % % % % % % % % % % % % % % % % % % % % % % % % % % % % % % % % % % % % % % % % % % % % % % % % % % % % % % % % % % % % % % % % % % % % % % % % % % % % % % % % % % % % % % % % % % % % % % % % % % % % % % % % % % % % % % % % % % % % % % % % % % % % 

% Scalar + Tensor

In the presence of NP, the effective Hamiltonian for the scattering process
$\nu_{\ell}+ n \to \ell^- + p$ 
  can be written in the form \cite{ccLag},
\bea
{\cal{H}}_{\rm eff} &=& \frac{4 G_F V_{ud}}{\sqrt{2}} \Big[ (1 + V_L)\,[\bar{u} \gamma^\mu P_L d] ~ [\bar{l} \gamma_\mu P_L \nu_l] \, +  V_R \, [\bar{u} \gamma^\mu P_R d] ~ [\bar{l} \gamma_\mu P_L \nu_l] \nl && \, + S_L \, [\bar{u} P_L d] \,[\bar{l}  P_L \nu_l] \, +  S_R \, ~[\bar{u} P_R d] \,~ [\bar{l}  P_L \nu_l]  \,  + T_L \, [\bar{u} \sigma^{\mu \nu} P_L d] \,~[\bar{l} \sigma_{\mu \nu} P_L \nu_l]\Big]\,,
\label{Heff}
\eea 
where  $G_F = 1.1663787(6) \times 10^{-5} GeV^{-2}$ is the Fermi coupling constant, $V_{ud}$ is the Cabibbo-Kobayashi-Maskawa (CKM) matrix element, $P_{L,R} = ( 1 \mp \gamma_5)/2$  are the projectors of negative/positive chiralities. We assume the neutrino to be always left chiral. The Hamiltonian can be written as,
\bea
{\cal {H}}_{\rm eff}&=&\frac{G_F V_{ud}}{\sqrt{2}}\left\lbrace \left[ \bar{u}\gamma^\mu(1 - \gamma_5)d\right]\; \left[\bar{l}\gamma_\mu (1-\gamma_5)\nu_{l}\right]+ \left[ \bar{u}(A_S +B_S \gamma_5)d\right]\; \left[\bar{l} (1-\gamma_5)\nu_{l}\right] \right.\nonumber\\
& +&\left. \left[ \bar{u}\gamma^\mu(A_V +B_V \gamma_5)d\right]\; \left[\bar{l}\gamma_\mu (1-\gamma_5)\nu_{l}\right]+ T_L \; \left[ \bar{u}\sigma^{\mu\nu}(1- \gamma_5)d \right]\; \left[\bar{l} \sigma_{\mu\nu}(1-\gamma_5)\nu_{l}\right] \right\rbrace,
\label{ST}
\eea
where $A_S=S_R+S_L$, $B_S=S_R-S_L$, $A_V=V_R+V_L$ and $B_V=V_R-V_L$ with $S_L$ and $S_R$ are the left and right handed scalar couplings, $V_L$ and $V_R$ are the left and right handed vector couplings and $T_L$ is the tensor coupling. The operators that describe the process $\bar{\nu}_{\ell}+ p \to \ell^+ + n$ can be obtained from the hermitian conjugate of the above Hamiltonian.
The co-effecients in the effective Hamiltonian are fixed by low energy observables such as $\tau$ decays \cite{Rashed:2012bd, Rashed:2013dba, Liu:2015rqa} .
%We first employ a model independent approach and treat the scalar and tensor couplings one at a time. 

 In $\scat$ the hadronic effects are described in terms of form factors. We
  define the charged hadronic current for the process $\scat$ in the SM as
\bea
\label{hadcureq3}
\bra {p(p^\prime)}J^+_\mu |n(p) \ket &=&   \bra {p(p^\prime)}(V_\mu -A_\mu)| n(p) \ket \nonumber\\
&=& V_{ud}\; \bar{p}(p^\prime) \Gamma_\mu n(p),
\eea
with
\bea
\Gamma_\mu &=& \left[  F^V_1 (t) \gamma_\mu + F^V_2 (t) i \frac{\sigma_{\mu\nu} q^\nu}{2M} + F_A (t) \gamma_\mu \gamma_5 + F_P (t) \gamma_5 \frac{q_\mu}{M} \right].
\eea
Here $F's$ are the hadronic form factors which are functions of the squared momentum transfer $t$.
The expressions for the vector and axial-vector hadronic currents in Eq.~\ref{hadcureq3} are
\bea
\label{vecAxME}
\bra {p(p^\prime)}V_\mu| n(p) \ket & =&V_{ud}\; \bar{p}(p^\prime) \Big[\gamma_\mu F^V_1 (t) + \frac{i}{2 M} \sigma_{\mu \nu} q^\nu F^V_2  (t)   \Big] n(p),\nonumber\\
-\bra {p(p^\prime)}A_\mu | n(p) \ket   &=&V_{ud}\; \bar{p}(p^\prime) \Big[\gamma_\mu  F_A (t) + \frac{q_\mu}{ M}  F_P (t) \Big]\gamma_5 n(p).
\eea

Similarly, in the presence of $(A_V, B_V)$,
\bea
\label{hadcureq33}
\bra {p(p^\prime)}J'^+_\mu |n(p) \ket &=&   \bra {p(p^\prime)}(A_V V_\mu +B_V A_\mu)| n(p) \ket,
\eea
with 
\bea
\label{vecAxMEE}
\bra {p(p^\prime)}A_V V_\mu| n(p) \ket & =&V_{ud}\; A_V \bar{p}(p^\prime) \Big[\gamma_\mu F^V_1 (t) + \frac{i}{2 M} \sigma_{\mu \nu} q^\nu F^V_2  (t)   \Big] n(p),\nonumber\\
\bra {p(p^\prime)}B_V A_\mu | n(p) \ket   &=&-V_{ud}\; B_V \bar{p}(p^\prime) \Big[\gamma_\mu  F_A (t) + \frac{q_\mu}{ M}  F_P (t) \Big]\gamma_5 n(p).
\eea

The scalar current for the process $\scat$ can be parametrized as follows
\bea
\label{hadcureq}
\langle p(p^\prime)| J^+ |n(p) \rangle &=&  \langle p(p^\prime)|\bar{u}(A_S + B_S \gamma_5)d| n(p)\rangle \nonumber\\ 
&=& V_{ud} \; \bar{p}(p^\prime) (A_S G_S + B_S G_P \gamma_5) n(p).
 \eea
Using the equation of motion, 
\bea
G_S (t) &=&  r_N  F^V_1(t),\;\; \mbox{with}\;\;  r_N  =\frac{M_n-M_p}{m_d-m_u}\sim  {\cal{O}} (1), \nonumber\\
%G_P (t)  &=& - \frac{M}{\bar{m}_q}[ F_A(t) + 2 x_t  F_P(t)],
G_P (t) &=& - \left(F_A(t)\left(\frac{M_n-M_p}{m_d-m_u}\right)+F_P(t)\frac{m_d+m_u}{M}\right),
\label{formfactor1}
\eea
with $m_u=2.3$ MeV and $m_d = 4.8$ MeV \cite{Agashe:2014kda}. 

In the presence of tensor state, the tensor current can be parametrized as follows
\bea
\label{hadcureq23}
\langle p(p^\prime)| J^{\mu\nu} |n(p) \rangle &=& \langle p(p^\prime)|\bar{u}\sigma^{\mu\nu}(1- \gamma_5)d| n(p)\rangle \nonumber\\ 
&=&iV_{ud} \; K_{S,P}\; \bar{p}(p^\prime) (\Gamma^\mu \tilde{\Gamma}^\nu-\tilde{\Gamma}^\mu \Gamma^\nu) n(p)\nonumber\\
&=&\frac{i}{2M} V_{ud} \; \bar{p}(p^\prime) (K_S \Pi_1^{\mu\nu}-K_P \Pi_2^{\mu\nu}\gamma_5) n(p),
 \eea
with $\tilde{\Gamma}$ defined as 
\bea
\tilde{\Gamma}^\mu (p,p') &=& \gamma_0 \Gamma^{\mu^\dagger} (p',p)\gamma_0 , \nonumber\\
&=&\left[  F^V_1 (t) \gamma_\mu - F^V_2 (t) i \frac{\sigma_{\mu\nu} q^\nu}{2M} + F_A (t) \gamma_\mu \gamma_5 - F_P (t) \gamma_5 \frac{q_\mu}{M} \right],
\label{Gamma}
\eea 
and
\bea
K_S &=& - \frac{M}{4t}\frac{(M_p^2 - M_n^2)-(m_u^2 - m_d^2)}{(M_p - M_n)}\frac{G_S}{F_A F_P},\nonumber\\
K_P &=& - \frac{M}{4t}\frac{(M_p^2 - M_n^2)- (m_u^2 - m_d^2)}{(M_p + M_n)}\frac{G_P}{F_1 F_P},
\eea
with
\bea
\Pi_1^{\mu\nu} &=& F_1 F_2 (\gamma^\mu \gamma^\nu \slashed q - 2 \gamma^\mu \slashed q \gamma^\nu + \slashed q \gamma^\mu \gamma^\nu  ) - 4 F_A F_P (\gamma^\nu q^\mu + \gamma^\mu q^\nu),\nonumber\\
\Pi_2^{\mu\nu} &=& F_A F_2 (\gamma^\mu \gamma^\nu \slashed q - 2 \gamma^\mu \slashed q \gamma^\nu + \slashed q \gamma^\mu \gamma^\nu  ) - 4 F_1 F_P (\gamma^\nu q^\mu + \gamma^\mu q^\nu). 
\eea

The total differential cross section is
\bea
\frac{d \sigma_{tot} (\nu)}{dt} &=& \frac{G_F^2 \; cos^2 \;\theta_c}{32 \pi  E_\nu^2 M^2}\left[ A_{tot} + B_{tot}\; (s-u) + C_{tot}\; (s-u)^2 \right], 
\eea
with
\bea
A_{tot} &=&  16  M^4 (x_t-x_l) \left[A_{V\pm A}+ A_{Sr} + A_{T}  + A_{V\pm A - Sr}+ A_{V\pm A - T} \right],     \nonumber\\
B_{tot}&=& 8  M^2  \left[B_{V\pm A} + B_{V\pm A -Sr}+B_{V\pm A -T}+B_{Sr-T}+B_T\right],\nonumber\\
C_{tot}&=&  C_{V\pm A} +C_{T} ,
\eea
where
\bea
A_{V\pm A} &=&(1+A_V)^2 \left[ F_1^2 \left(1+x_l+x_t\right)+F_2^2 (x_l+ x_t^2 + x_t)+ 2 F_1 F_2 \left( x_l+2 x_t\right) \right]     \nonumber\\
&+&(1-B_V)^2 \left[ F_A^2 (-1+x_l+x_t)+ 4 F_P^2  x_l x_t +  4 F_A F_P x_l  \right],\nonumber\\
A_{Sr} &=& A_S^2 G_S^2  (x_t-1)+B_S^2 G_P^2  x_t  ,\nonumber\\
A_T &=& 64 T_L^2 F_2^2 \left( F_1^2 K_S^2 (x_t -1)(x_l +x_t)+F_A^2 K_P^2   \left( x_l x_t +x_l+x_t^2\right)\right), \nonumber\\
A_{V\pm A - Sr} &=& -2(1-B_V)  B_S G_P  \sqrt{ x_l} (F_A+2F_P x_t)\left(1-4 x_t \frac{M^2}{M_W^2}\right),\nonumber\\
A_{V\pm A - T}&=& - 32 (1-B_V) T_L K_S F_1 F_2   F_A   \sqrt{x_l}(x_t-1) + 16 (1+A_V)F_2 F_A K_P T_L \sqrt{x_l} (2F_1 x_t + F_1+3F_2 x_t),\nonumber\\
\eea

\bea
B_{V\pm A} &=& 2 (1+A_V)(1-B_V) x_t F_A (F_1+F_2),\nonumber\\
B_{V\pm A - Sr} &=& (1+A_V) A_S G_S  \sqrt{x_l} (F_1+F_2 x_t) ,\nonumber\\
B_{V\pm A - T} &=&-16 (1+A_V) T_L K_S  x_t \sqrt{x_l} F_1 F_2 (F_1+F_2)+16(1-B_V)T_L K_P x_t \sqrt{x_l}F_2 F_A^2, \nonumber\\
B_{Sr-T}&=& 8 A_S G_S F_2 x_t F_A K_P T_L,\nonumber\\
B_T &=& -128 T_L^2 x_t x_l F_1 F_2^2 F_A K_S K_P,
\eea
\bea
C_{V\pm A} &=& (1+A_V)^2 (F_1^2-x_t F_2^2)+(1-B_V)^2 F_A^2,\nonumber\\
C_{T} &=& -64  T_L^2  F_2^2  x_t  (F_A^2 K_P^2+F_1^2 K_S^2).
\eea
Here, V$\pm$A stands for the SM and $(A_V, B_V)$ contributions, $Sr$ for scalar, and $T$ for tensor. $x_t = t/4M^2$ and $x_l=m_\tau^2 / 4 M^2$ where $M$ and $m_\tau$ are the nucleon and tau masses. $s, u, t $ are the Mandelstam variables

{}For the sake of simplicity we will consider two new physics scenarios. In one we have $S+T$ interactions and in the other $V \pm A$ interecations. Leptoquark models for instance produce  both $S$ and $T$ interactions while models with extra gauge bosons have $V $ and $A$ interactions.

The quasielastic scattering of an antineutrino on a free nucleon is given by 
\beq
\label{quasisceq111}
\bar{\nu}_\tau(k) + p(p) \to \tau^+(k^\prime) + n(p^\prime)\,.
\eeq
The charged hadronic current in SM becomes \cite{Hagiwara:2003di, Llewellyn Smith:1971zm}
  \bea
 \label{hadcureq000}
\langle n(p^\prime)|J^-_\mu |p(p) \rangle &=& \langle p(p)|J^+_\mu |n(p^\prime) \rangle^\dagger \nonumber\\
&=& V_{ud}\;  \bar{n}(p^\prime)\; \tilde{\Gamma}_\mu\; p(p),
 \eea
where
%\beq
%\tilde{\Gamma}_\mu (p,p')= \gamma_0 \Gamma_\mu^\dagger (p',p)\gamma_0 ,
%\eeq
%and  
$\tilde{\Gamma}_\mu$ is defined in Eq.~\ref{Gamma}.
%\bea
%\tilde{\Gamma}^\mu &=& \left[  F^V_1 (t) \gamma_\mu - F^V_2 (t) i \frac{\sigma_{\mu\nu} q^\nu}{2M} + F_A (t) \gamma_\mu \gamma_5 - F_P (t) \gamma_5 \frac{q_\mu}{M} \right].
%\label{Gamma}
%\eea

By comparing the processes $\scat$ and $\scatanti$, the expression of the differential cross section for anti-neutrino scattering can be obtained from the one for neutrino interaction by making the following changes: ${F_P \rightarrow -F_P,\; F_2^V \rightarrow -F_2^V,\; (s-u) \rightarrow -(s-u)}$, $M_p \leftrightarrow M_n$, and $m_u \leftrightarrow m_d$.

The scalar and tensor currents for the process $\scatanti$ are parametrized as 
\bea
\label{hadcureq32}
\langle n(p^\prime)| J^- |p(p) \rangle &=&  \langle n(p^\prime)|\bar{d}(A_S - B_S \gamma_5)u| p(p)\rangle \nonumber\\ 
&=& V_{ud} \; \bar{n}(p^\prime) (A_S \bar{G}_S - B_S \bar{G}_P \gamma_5) p(p).
 \eea
and
\bea
\label{hadcureq2}
\langle n(p^\prime)| J^{(-)\mu\nu} |p(p) \rangle &=& \langle n(p^\prime)|\bar{d}\sigma^{\mu\nu}(1+\gamma_5)u| p(p)\rangle \nonumber\\ 
&=&iV_{ud} \; \bar{K}_{S,P}\; \bar{n}(p^\prime) (\Gamma^\mu \tilde{\Gamma}^\nu-\tilde{\Gamma}^\mu \Gamma^\nu) p(p)\nonumber\\
&=&\frac{i}{2M} V_{ud} \; \bar{n}(p^\prime) (\bar{K}_S \Pi_1^{\mu\nu}+\bar{K}_P \Pi_2^{\mu\nu}\gamma_5) p(p),
 \eea

\begin{figure*}
\centering
\includegraphics[width=7cm]{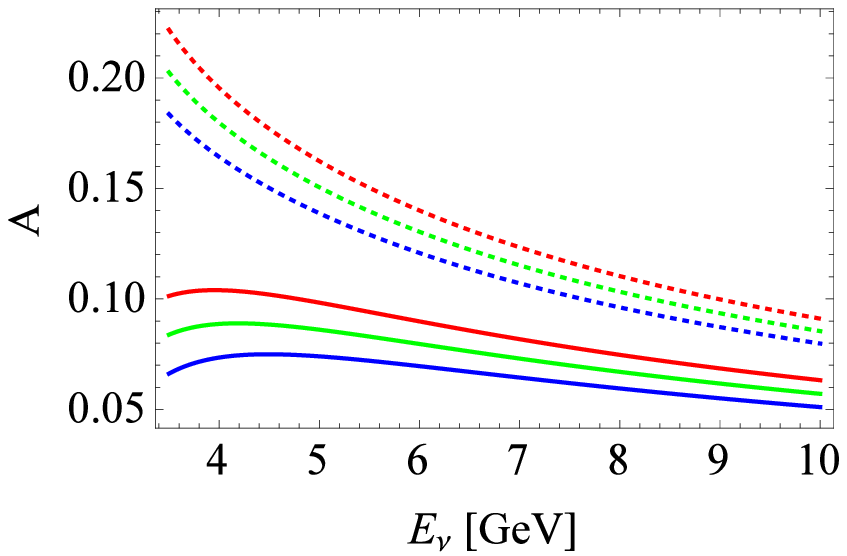}~~~
\includegraphics[width=7cm]{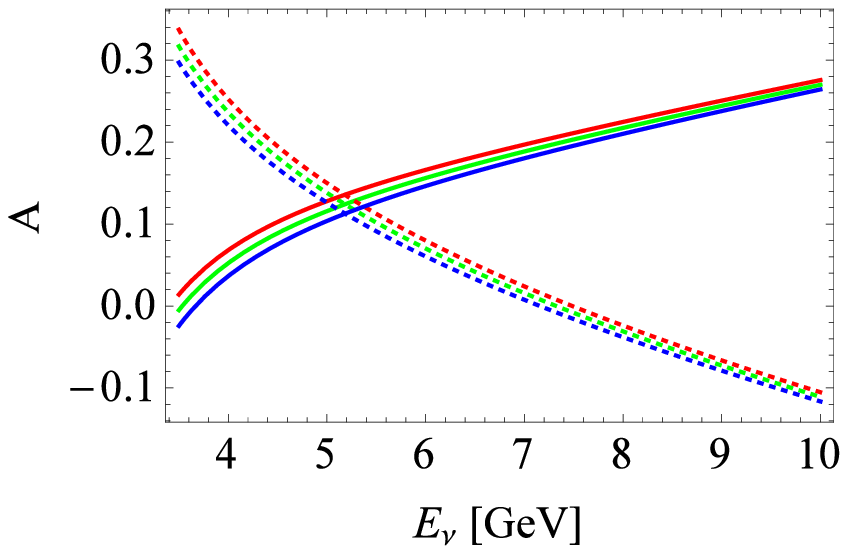}~~~\\
\includegraphics[width=7cm]{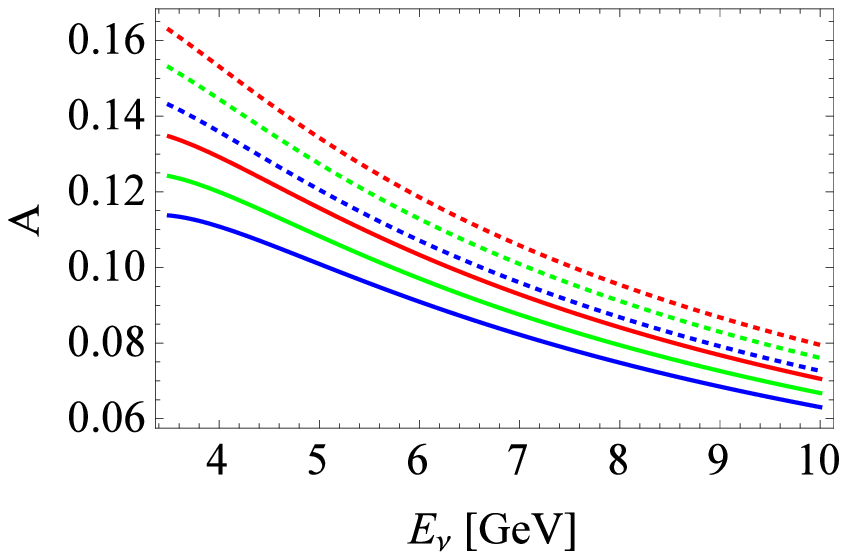}~~~
\includegraphics[width=7cm]{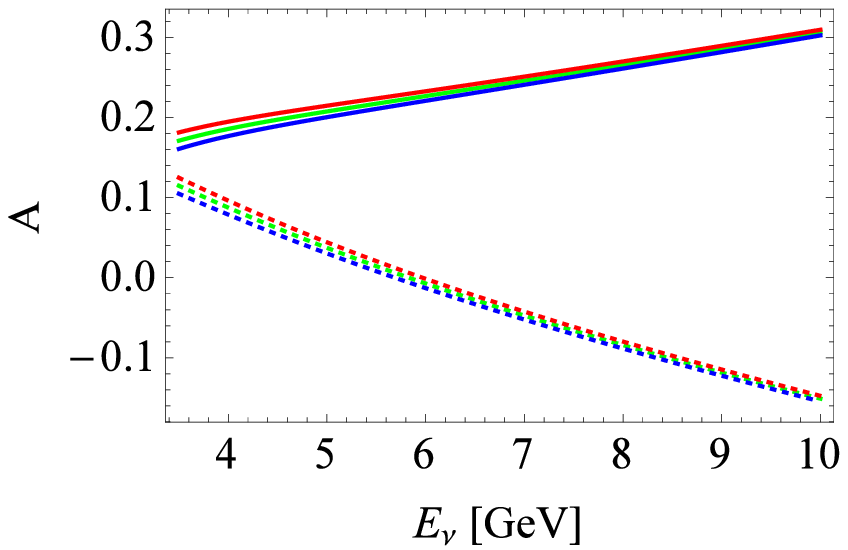}~~~
\caption{$S+T$ model:  The asymmetry parameter $A$ for the energy range and baseline relevant to LBNO (top) and DUNE (bottom) experiments, where quasielastic effect is dominant. 
The solid/dotted lines correspond to NH/IH. The green, red,  and blue lines correspond to $\delta=(0,\pi/2,-\pi/2)$, respectively,  with $(S_R, S_L, T_L)=(-1.87, -1.31, 0.18)$. Left Panel: NSI parameters are taken to be $(\varepsilon_{\mu\tau},\varepsilon_{\tau\tau})=(0,0)$. Right Panel: NSI parameters are taken to be $(\varepsilon_{\mu\tau},\varepsilon_{\tau\tau})=(0.07,0.147)$.}
\label{Figure2-6}
\end{figure*}

The form factors become 
\bea
\bar{G}_S (t) &=&  r_N  F^V_1(t),\;\; \mbox{with}\;\;  r_N  =\frac{M_n-M_p}{m_d-m_u}\sim  {\cal{O}} (1), \nonumber\\
%G_P (t)  &=& - \frac{M}{\bar{m}_q}[ F_A(t) - 2 x_t  F_P(t)],
\bar{G}_P (t) &=& - \left(F_A(t)\left(\frac{M_n-M_p}{m_d-m_u}\right)-F_P(t)\frac{m_d+m_u}{M}\right),
\label{formfactor11}
\eea
and
\bea
\bar{K}_S &=& - \frac{M}{4t}\frac{(M_n^2 - M_p^2)-(m_d^2 - m_u^2)}{(M_n - M_p)}\frac{\bar{G}_S}{F_A F_P},\nonumber\\
\bar{K}_P &=& - \frac{M}{4t}\frac{(M_n^2 - M_p^2)-(m_d^2 - m_u^2)}{(M_n + M_p)}\frac{\bar{G}_P}{F_1 F_P}.
\eea
The effect of the scalar-tensor interactions at detection, on the asymmetry parameter $A$ is shown in Fig.~\ref{Figure2-6}. We can make the following observation:

\begin{itemize}

\item{ With no NSI along propagation there can be  difference for the $A$ parameter between the two hierarchies though it is always positive. This difference in $A$ is more appreciable for the LBNO baseline compared to the DUNE experiment.}

\item{ When NSI along propagation is included large differences in the $A$ parameter is possible for both the baselines specially at larger energies. {} For both baselines the $A$ parameter increases with energy for NH and decreases with energy for the IH. For the LBNO baseline, the crossing point when no NSI at detection is considered happens at $E_\nu=5$ GeV and $A=0$. When $S+T$ contribution is included, the crossing point remains at the same energy value but at non-zero value of $A$. This means that a non-zero value of $A$ can be observed, but still the MH  would not be resolved.}

\end{itemize} 

\begin{figure*}
\centering
\includegraphics[width=7cm]{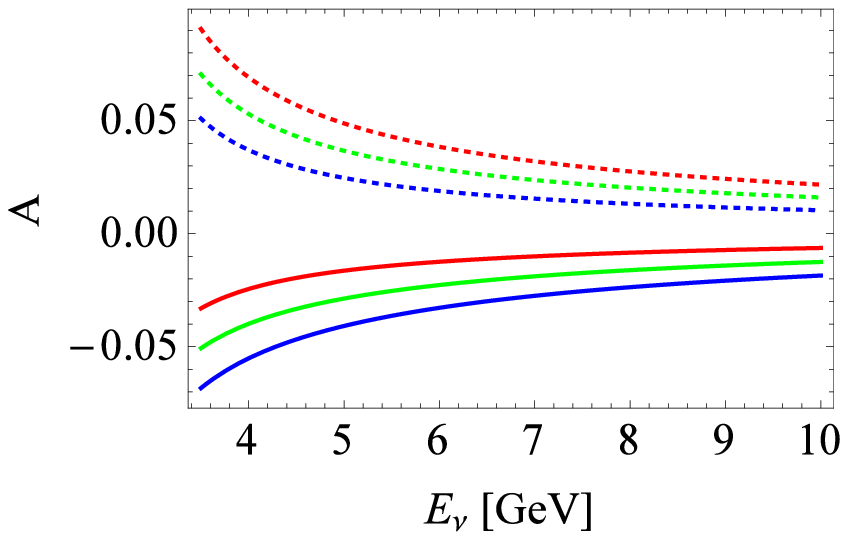}~~~
\includegraphics[width=7cm]{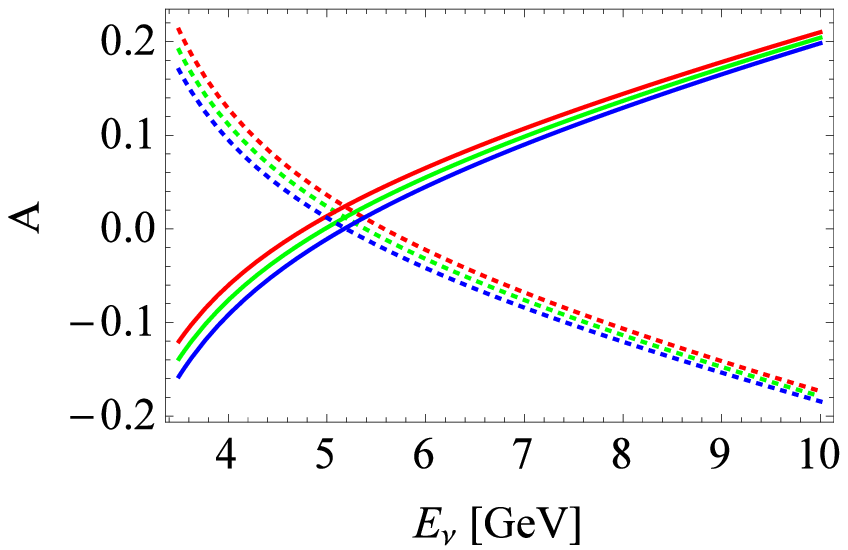}~~~\\
\includegraphics[width=7cm]{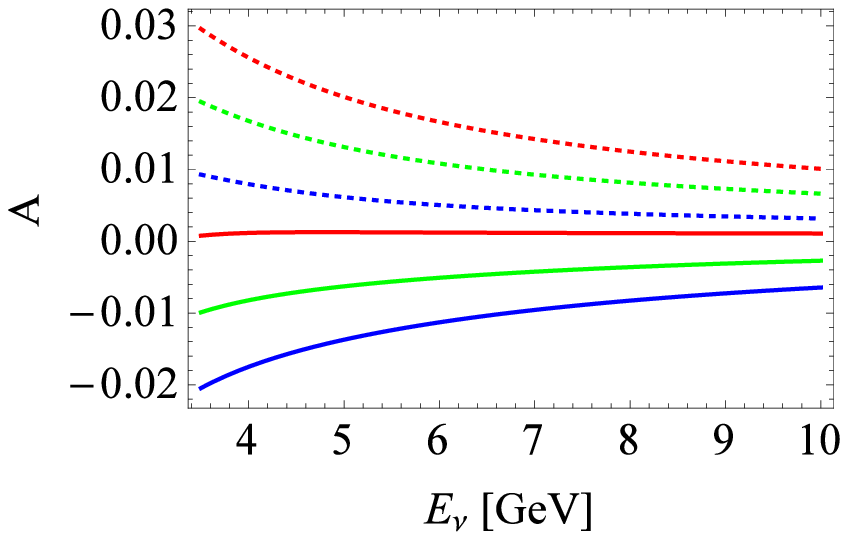}~~~
\includegraphics[width=7cm]{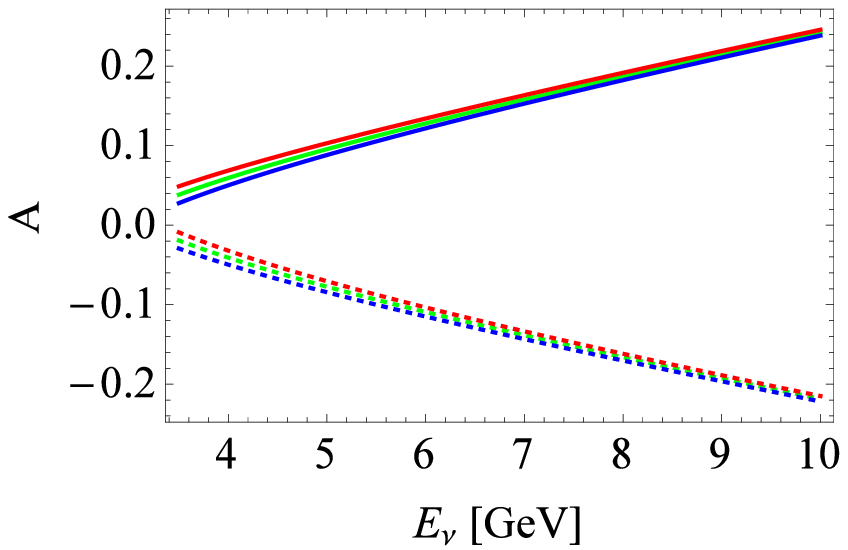}~~~
\caption{V$\pm$A model: The asymmetry parameter $A$ for the  energy range and baseline relevant to LBNO (top) and DUNE (bottom) experiments, where quasielastic effect is dominant. 
The solid/dotted lines correspond to NH/IH. The green, red,  and blue lines correspond to $\delta=(0,\pi/2,-\pi/2)$, respectively, with $(V_L, V_R)=(0.016,0.006
)$ . Left Panel: NSI parameters are taken to be $(\varepsilon_{\mu\tau},\varepsilon_{\tau\tau})=(0,0)$. Right Panel: NSI parameters are taken to be $(\varepsilon_{\mu\tau},\varepsilon_{\tau\tau})=(0.07,0.147)$.}
\label{Figure2-4}
\end{figure*}

The effect of the V$\pm$A interactions at detection, on the asymmetry parameter $A$ is shown in Fig.~\ref{Figure2-4}. The $V \pm A$ interactions are more tightly constrained than the $ S+T$ models and so in this case the effect of NSI at detection on the $A$ parameter is modest and the general features of the $A$ parameter does not alter significantly when compared with no NSI at detection.

% % % % % % % % % % % % % % % % % % % % % % % % % % % % % % % % % % % % % % % % % % % % % % % % % % % % % % % % % % % % % % % % % % % % % % % % % % % % % % % % % % % % % % % % % % % % % % % % % % % % % % % % % % % % % % % % % % % % % % % % % % % % % % % % % % % % % % % % % % 

%\section{Conclusion}

In conclusion,  in this work we explored  resolving the MH using
the $\mutau$ and $\mutaubar$ appearance channels. 
The determination of the mass hierarchy in these channels has an advantage over $\nu_\mu \rightarrow \nu_e$ as $P(\nu_\mu \rightarrow \nu_\tau)$ is not suppressed by small oscillation parameters as in the case for $P(\nu_\mu \rightarrow \nu_e)$.
We also considered NSI effects along propagation and at detection for these channels.These transitions
 can be accessible with good precision  for baseline as in the LBNO experiment. The DUNE experiment will have  access to these channels but with less number of events.
 %Sufficient number of events in LBNO will be observed, that is, the transition probability of these two channels can be generated with high reliability.
 To resolve the MH, we introduced an asymmetry parameter defined as the difference of the two probabilities $\Pmutau$ and $\Pmutaubar$ normalized to their sum. The energy range of the LBNO and DUNE experiments is $(0-10)$ GeV, where the quasielastic (QE) scattering is dominant. We found that in this energy region the asymmetry parameter is positive for inverted hierarchy and negative for normal hierarchy when  NSI are ignored.  The prospect for resolving the hierarchy is much better for a LBNO type baseline. But when NSI along propagation is considered, the sign of $A$  is flipped as we vary the energy.  In both LBNO and DUNE baselines appreciable differences between the $A$ parameters for the two hierarchies can arise. This gives a clear signal of the mass hierarchy as well as NSI.  The pattern of the asymmetry parameter was found to be modestly sensitive to the CP phase.  
 %Because of the hints of the non-universality in the lepton sector emerged in the BaBar and LHCb experiments, we assume the largest new physics %effects are in the tau sector. We considered the NSI parameters $(\varepsilon_{\mu\tau},\varepsilon_{\tau\tau})$ along propagation, 
 We considered NSI at detection in an effective Hamiltonian framework with generic vector axial-vector,  scalar, and tensor interactions. The parameters in the effective Hamiltonian are constrained by $\tau$ decays. We found the $S+T$ model could have significant impact on the $A$ parameter though the MH could still be resolved in a LBNO type baseline and in DUNE if NSI in propagation is present.
%$\Pmutau$ is insensitive to $\delta_{\rm CP}$ which means that the main source of  distinction between $\mutau$ and $\mutaubar$ comes only from the matter effect. This makes the measurement of the mass hierarchy unique.

%\pagebreak
% % % % % % % % % % % % % % % % % % % % % % % % % % % % % % % % % % % % % % % % % % % % % % % % % % % % % % % % % % % % % % % % % % % % % % % % % % % % % % % % % % % % % % % % % % % % % % % % % % % % % % % % % % % % % % % % % % % % % % % % % % % % % % % % % % % % % % % % % % 

%\section*{Acknowledgements} 

\textit{Acknowledgements:} A.R acknowledges the hospitality of ICTP, Trieste, Italy during a visit when this work was in progress. This work was financially supported in part by the National Science Foundation under Grant No.NSF PHY-1414345.

\end{document}